\documentclass[12pt]{article}
\usepackage{graphicx}

\begin{document}

\title{Magnetism induced by electric impurities: \\
A one-body problem with half the physics of
the fractional quantum Hall effect}

\author{Alfred Scharff Goldhaber$^*$ and M. L. Horner$^\dagger$ \\
\small{$^*$C.N. Yang Institute for Theoretical Physics}\\
\small{State University of New York}\\ 
\small{Stony Brook, NY 11794-3840}\\
\small{and}\\
\small{$^\dagger$Department of Physics}\\ 
\small{Southern Illinois University}\\ 
\small{Edwardsville, IL 62026}}

\date{\today}

\maketitle

\begin{abstract}

We study spectra for a 2D electron in the lowest Landau level with
randomly distributed, repulsively correlated electric impurities.  The
 lowest energy band reflects an effective magnetic field
downshifted  by an integer multiple of  the impurity
density.  The  downshift is precisely half that for the corresponding
electron density in the composite-fermion picture of the fractional
quantum Hall regime. 
\end{abstract}

PACS number(s):  73.20.At, 73.43.-f

\section{Introduction}

In a famous paper on the relation between mathematics and physics  \cite{Wigner60},
Eugene Wigner speculated on the
possibility of phenomena that cannot be built from something `more
fundamental'.  Within physics, reduction
to fundamentals has generally taken the form of going to smaller distances and
 higher energies, hoping to find fewer and simpler rules. Phenomena
at large distances or low energies, even some once regarded as fundamental,
then become consequences of the small-distance or high-energy behavior.
In practice, even though
experimental and numerical evidence  may give a link great conviction,  rigorous proofs are quite rare.  This
is one reason why experiment is essential to the progress of physics, and 
also why successful prediction of `less fundamental' physics also is rare.

 Anderson \cite{pl} and Laughlin \cite{rbl} introduced an appealing perspective on such questions, defining ``emergent phenomena" that catch physicists by surprise, even 
though they might in principle have been deduced from more microscopic knowledge.  Their articles  emphasize that such occurrences are associated especially with systems including large numbers of components, leading to the phrase ``more is different."

The fractional quantum Hall effect (FQHE) is a prime example of such
unpredicted physics.  Following the experimental  discovery,  
 Laughlin \cite{Laughlin83} gave an explanation, providing an ansatz for the many-body
wave functions relevant to  the simple FQHE fractions. 
In a further advance Jain \cite{Jain89a, Jain89b,
Jain90, Jain97} introduced `composite fermions' to account for general fractions.
The CF description works well even for the compressible regime where the ratio of electron density
to magnetic flux density is around
$\nu=\frac{1}{2}$.  The evidence justifying these approaches comes from experiment, from exact
numerical computations for small systems, and from self-consistent
computations for larger systems. 

These successes leave little doubt that we have a valid theory for FQHE,
but it is tempting to ask for more:
One may assume that only
electrons in the lowest Landau level need be considered, very nearly
reducing a problem in two space dimensions to a problem in one space
dimension.  Thus this might be one of those rare cases where rigorous 
deduction of the phenomena from the (very simple) fundamental dynamics
is within reach.

We aim to establish
a pathway towards understanding the FQHE 
as a deducible phenomenon.
In terms of one quantitative measure, we shall see that the first
steps described below come exactly halfway towards that goal,
although more by numerical than by analytic methods. 
The starting point should be a two-dimensional
gas of electrons whose screened electric interaction gives a short-range
repulsion, with a magnetic field in the direction normal to the plane
of electron motion so strong that only the lowest Landau level for
one orientation of electron spin need be considered.

In this work we make a drastic additional simplification, by treating all
electrons except one as fixed in position, and consider the resulting
one-body problem of an electron moving in a magnetic field under the
influence of an array of electric impurities mimicking the effect of
all the other electrons.  Because these electrons have strong repulsive
correlations, we assume such correlations for the impurities also. 
Studying a system with tens to thousands of impurities spread throughout
a region of uniform magnetic field, we find that the one-body spectrum
shows a characteristic level clustering which implies an effective
magnetic field shifted down from the applied field by an integer number
$n$ of flux quanta per impurity.  In Jain's composite fermion model, a
field shift specified by an even integer $2q$ is found, where the value
of $q$ increases as the impurity density decreases.  We find the
remarkable result for densities in a range from  $\nu=\frac{1}{3}$ towards $\nu=\frac19$,
$q=n$, i.e., the two integers track together.  Thus by this measure our one-body problem
accounts for exactly half the physics in the fractional quantum Hall
regime.

The crucial idea behind our study is that an electron in the presence of
electric impurities and a strong magnetic field feels an induced or effective
magnetic field transmuting the electric repulsion into magnetic torque.
To motivate this idea, consider a two-dimensional Aharonov atom
\cite{Aharonov} in a uniform, perpendicular magnetic field. Let   the
attractive potential binding the electron to the (uncharged) `nucleus'
 constrain the electron to occupy only degenerate states centered on
 the nucleus with azimuthal quantum numbers 
 $m=0$ or $m=1$ of the lowest Landau
level. The degeneracy will be broken by interactions of the atom
with electric impurities.
 Imagine that the nucleus is
allowed to move towards a concentrated charge at point ${\bf r}$.  For
large separations, the $m=0$ state of the atom will give the least
overlap of the electron and charge, and thus the lowest energy. 
However, at zero
separation the $m=1$ state of the atom will give the lowest energy.  

Because there
is  mixing between the two states, the atom will shift from the $m=0$
to the $m=1$ state in the course of an adiabatic journey from large to
zero separation between the nucleus and the fixed electron.  In the
limit of weak mixing the transition from $m=0$
to $m=1$ takes place on a circle specified by  a precise separation $R$ between the charge
and the nucleus of the Aharonov atom. Therefore, when the nucleus of the atom sits at
that separation, the two wavefunctions must match everywhere on
the transition circle. 
This can be achieved provided there is a gauge transformation
matching the ``atomic" wave functions 
across the transition circle, given 
by the phase factor $e^{i\phi}$. The result of this gauge
transformation is equivalent to an additional effective magnetic flux of one
quantum, opposing the uniform magnetic field.  Because the mixing by the Coulomb field is not infinitely weak, the
transition circle becomes a diffuse ring, in which the effective flux
is distributed.  
If the nucleus moves slowly in a closed curve, the  effective or induced magnetic flux enclosed is an example of
Berry's `geometric magnetism'  \cite{Berry:1984jv}.
Clearly, this effect could not occur if the given uniform
magnetic field were not present, because there would be no way for the
Coulomb repulsion to ``choose'' the sign of the effective magnetic
field.  

The fictitious Aharonov atom suggests a more realistic example.
Let us  choose as our slow variable,
in place of the heavy nucleus, the guiding-center coordinate for the
electron, allowing in principle the full (degenerate) set of azimuthal $m$ values with respect to this
center.  This would specify a complete set of
states for any location of the guiding center, but the requirement of
minimum energy in the presence of electric impurities should remove
the consequent redundancy.  

Now imagine some density of distributed electric impurities.
If the density is low enough, then when the guiding center is exactly in
a space between impurities, as far as possible from each of the nearest neighbors,
it is energetically favorable to have $m=0$.   On the other hand, if
the guiding center coincides with one of the impurities, a higher $m$ value
minimizes the energy, by keeping the electron away  from the central
impurity.   The new $m$ value cannot be too high, however, to avoid 
overlap with other impurities.  Thus the qualitative expectation is that
the optimum $m$ value for this position will increase as the impurity density
decreases.  The  consequence will be an effective magnetic
field, always of one sign, distributed throughout the plane, and on
average an integer multiple of the impurity density, with the multiple
being larger as the density becomes smaller. 

We shall see that such behavior indeed occurs for the lowest-energy
part of the spectrum in the presence of randomly distributed, 
repulsively correlated impurities.  It doesn't happen for a regular 
lattice, most likely because some higher $m$ values then can
be degenerate with the two specified in the above discussion.

Our approach superficially resembles earlier models for genuine impurities, related to the
integer quantum Hall effect. The main differences between this analysis and
those studies \cite{Aoki77, Aoki78, Ando83, Ando84a, Ando84b, 
Huo92, Huo93, Liu94} are: a) impurity density of less than one per flux quantum
for our analysis, rather than five or more per flux quantum for the
earlier studies, and b) our identification of the impurities as  other
electrons, distributed with repulsive correlations,  as opposed to uncorrelated  lattice and surface irregularities and atomic
impurities.  

\section{Methods}

To describe a single electron moving in two dimensions in a magnetic
field, we use the Hamiltonian
\begin{equation} \label{hamiltonian}
H = \frac{1}{2} \left[ {\bf \sigma } \cdot \left( {\bf p} - e {\bf A}
\right) \right]^2 ,
\end{equation}
in units where $\hbar=c=m=1$ and with $\bf \sigma_i$ the Pauli spin
matrices.  This Hamiltonian includes the interaction of the electron's
spin with the magnetic field, and implies that the lowest Landau level
states are spin polarized, with energy $E=0$.  We choose the cylindrical
gauge and consider the magnetic field B to be strong enough that we can
consider only the lowest Landau level and neglect mixing. In keeping
with the cylindrically symmetric gauge, we choose our basis states to be
concentric orthonormal eigenfunctions of the form \begin{equation}
\label{wavefunction}
\phi_m \left( z \right) = \frac{1}{\sqrt{\pi m !} } z^m e^{-zz^*/2}
\end{equation}
where $z=x+iy=re^{i\theta}$ has been expressed in units of the magnetic
length
\begin{equation} \label{magneticlength}
\ell_B = \sqrt{2/eB}.
\end{equation}
The magnetic length is both the length scale of the root-mean-square
radius of the $m=0$ basis state and the radius of the area through which
one quantum of magnetic flux, $\Phi_o=2\pi/e$, passes.  The total number
of flux quanta passing through the system is then given by the squared
radius of the system. That radius we take as the root-mean-square radius
of the largest basis state. We construct our potential out of sums of
impurities where each impurity is a pair of Gaussians,
\begin{equation}
\label{paired}
V_n \left( z\right) = V_0 \left[
\frac{1}{\alpha^2}
e^{-\left|z-z_n\right|^2/\alpha^2} -\frac{1}{\alpha^2\beta^2}
e^{-\left|z-z_n\right|^2/\alpha^2\beta^2} \right] \ \ .
\end{equation}
The use of gaussians or other
short-range potentials has been common in numerical FQHE studies,
e.g., \cite{Laughlin83}.
The paired Gaussians give a system with zero average electrical 
potential and therefore a total energy $E=0$, and approximate a 
repulsive charge screened by a compensating cloud.  For moderate 
impurity densities, we set $\beta^2=2$ so that the area computed from 
the root-mean-square radius of the negative shell would be twice that 
of the positive core, and then choose $\alpha$ so that plots of the density of 
states, participation ratio, and root-mean-square radius as a function of energy
are as 
symmetric  with respect to energy as possible for a density of one impurity for every two 
flux quanta. 
 The widths of the Gaussians then scale inversely with the square root of the 
impurity density, as we expect increased impurity density to 
increase localization of the individual impurities to minimize their 
overlap and hence their interaction energy.  The potentials are 
constructed so that their space integrals are unchanged by this 
scaling.  The impurities are randomly distributed, or randomly 
distributed with a minimum separation between adjacent impurity 
centers (hard-shell constraint), or placed on a hexagonal lattice. 
The centers of the impurities are required to lie within the disk, 
but their shape is unaffected by the disk boundary.  The impurity 
density $\nu$ is the number of impurities per flux quantum.

\section{Results}
We consider energy spectra and compactness, ${\cal C}$, of states.  The latter is defined by 
 \begin{equation}
{\cal C}=\sqrt{P}/R_{rms}.\label{compactness}
\end{equation}
The compactness measure uses the participation ratio, which measures the area of a state,
\cite{Bell70,Edwards72,Thouless74}, 
\begin{equation}
P_{\alpha}=\frac{\left(\int d^{2}r\left|\Psi_{\alpha}\right|^{2}\right)^{2}}{\int d^{2}r\left|\Psi_{\alpha}\right|^{4}}
=\left(\int d^{2}r\left|\Psi_{\alpha}\right|^{4}\right)^{-1}.\label{participation}
\end{equation}
and  the root-mean-square radius, which measures
the width of the state, 
\begin{equation}
R_{rms}^{\left(\alpha\right)}=\sqrt{\int d^{2}rr^{2}\left|\Psi_{\alpha}\right|^{2}}.\label{rms}
\end{equation}
Compactness measures the degree to which a state resembles a disk
or a ring and has an intrinsic scale of $\sqrt{2\pi}$. For a uniform
disk and densities corresponding to the $m=0$ and $m=1$ basis states,
the ratio has exactly the value $\sqrt{2\pi}$. This is not a maximum
value as can be seen by permitting $m$ to be a continuous rather than
a discrete variable. In the limit of a ring of vanishing thickness,
the value of the ratio tends toward zero. 

\begin{figure}
\begin{center}
\includegraphics[scale=1]{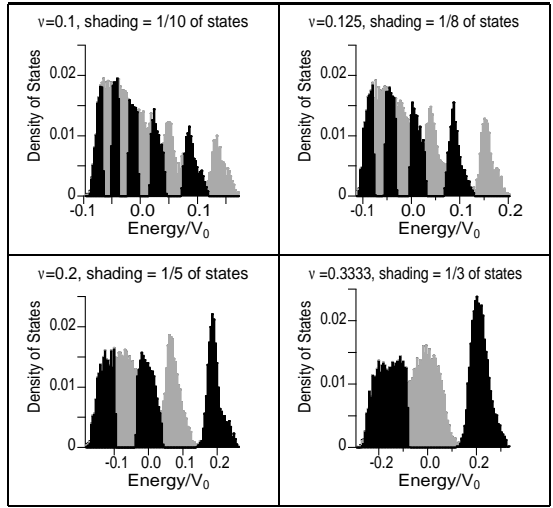}
\caption{\small
{Summed energy histograms for ten different random
distributions of impurities for the case of parameters set to  $\alpha^{2}=\frac{0.35}{\nu}$,
$\beta^{2}=2$, minimum distance between centers of impurities is
$0.70\frac{SystemSize}{\sqrt{\# Impurities}}$, using matrix elements
up to 201 places from the diagonal. The filling factor, $\nu$ is
indicated at the top of each subfigure. Shading groups states into
bands containing $\nu$ of the states counting from the highest energy.}}
\label{spectra} 
\end{center}
\end{figure}

\begin{figure}
\begin{center}
\includegraphics[scale=1]{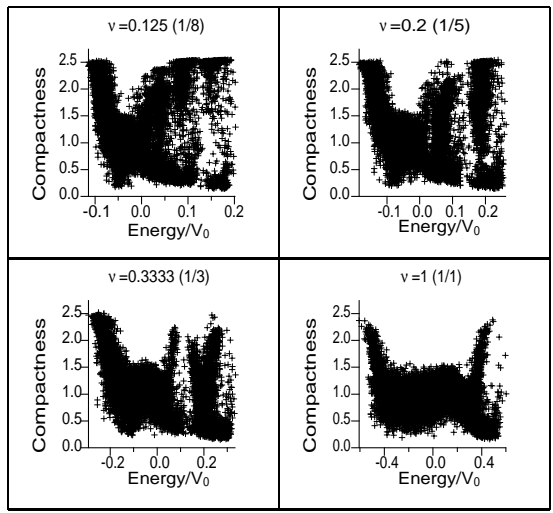}
\caption{ \small{Combined distributions of compactness as
a function of energy for ten different random distributions of impurities
for the case of parameters set to $\alpha^{2}=\frac{0.35}{\nu}$, $\beta^{2}=2$,
minimum distance between centers of impurities is $0.70\frac{SystemSize}{\sqrt{\# Impurities}}$,
using matrix elements up to 201 places from the diagonal and cross-terms
in the participation function no more than 100 apart. The filling
factor, $\nu$ is indicated at the top of each subfigure. }}
\label{shape}
\end{center}
\end{figure}

Examining the energy spectra for the different kinds of distributions at
impurity densities between 7 and 500 impurities per 1000 magnetic flux
quanta, we find that for double Gaussian impurities randomly
placed with a hard-shell constraint, the energy spectrum  consists of a fraction $1-n\nu$ of the
states in the lowest energy band, and $n$ higher energy bands each
containing a fraction $\nu$ of the states.  This occurs only for hard-shell repulsion with packing
fractions approaching hexagonal close packing.  Neither a random distribution nor a precise
hexagonal close packing produces this type of energy spectrum.  We also find that
compactness as a function of energy for the lowest energy band is qualitatively
similar to that for the single band that exists for one impurity per flux quantum.
Both of these observations suggest that the lowest band is a filled Landau level
in a reduced magnetic field.  According to \cite{Jain97},
when the filling factor is given by 
\begin{equation}
\nu=\frac{p}{2pq+1},\label{filling}
\end{equation}
 then the effective magnetic field is given by 
 \begin{equation}
B^{*}=B\left(1-2q\nu\right).\label{effective}
\end{equation}

For $\nu=\frac13$ and $\nu=\frac15$, we see our effective field shifted from the applied field by exactly
 half the amount found in the composite
fermion picture. At impurity densities falling between the
discrete electron filling factors permitted in the composite fermion
picture, there is a smooth transition in the numbers of upper bands
split off from the lowest band;  only a limited number of bands will
split off.  In the vicinity of $\nu=\frac19$, the upper bands gradually lose
their definition and are reabsorbed into the lowest energy band:  At $\nu=\frac18$ we have
$n=4$, consistent with the value of $q$ in CF theory for $\nu=\frac19$, while for $\nu=\frac1{10}$ all the
bands have merged. 

 The
appearance and the disappearance of banding with decreasing impurity density
implies the presence of an effective magnetic field which eventually
dissolves  at still lower impurity densities.  Comparison with single
Gaussians and unconstrained random and crystalline distributions shows
that the separation induced by the combination of the double Gaussian
and the hard-shell constraint is critical to the phenomena described
above.  This
tells us that the energy banding is linked to a moderate amount of order
in the system - neither too much nor too little should be present.

\section{Conclusions}

We find that
in the case of hard-shell repulsion
 the number and character of the single-electron states in the
lowest band correspond to the  lowest Landau level in an effective magnetic field. Interpreting
the impurities as the other electrons of a fractional quantum Hall
system, the effective field is shifted down from the applied field by one
half the amount   expected from CF theory.  Seeing the effect only for hard-shell repulsion
makes sense, because this case best represents the known behavior of electrons in the FQHE.
The fact that the bands are  diffuse, rather than sharp as one would expect in a complete
theory, is inevitable for fixed impurities.

Our strongest constraint is the restriction to just one dynamical electron,
  clearly
 losing much of the electron-electron dynamics crucial
for the FQHE.  This means that we give up any representation of Fermi-Dirac statistics, surely crucial to a complete theory.
Partially relaxing the one-particle constraint, by considering simultaneously motion of each of a pair of electrons (in the spirit of Cooper's consideration \cite{coop} of an electron pair in the background of an electron Fermi sea as a way of understanding superconductivity), might perhaps give the full reduction of
the magnetic field seen in  CF theory, with wave-function
antisymmetrization  selecting odd-denominator rather than
 even-denominator filling factors.  In any case, completing the project of deducing FQHE still would require treating all the electrons as dynamical, with the guiding-center coordinates as the adiabatic variables.  

\section*{Acknowledgments}
Parts of this paper are based on work in partial fulfilment of the
requirements of a Ph.D. at  Stony Brook University for MLH.  Partial funding
for this research came from the Creighton University 
Physics Department, and from the National Science Foundation,
Grant PHY-0140192.

\end{document}